\documentclass[english,aps,prl,preprint]{revtex4-1}
\usepackage[T1]{fontenc}
\usepackage[latin9]{inputenc}
\usepackage{textcomp}
\usepackage{amsmath}
\usepackage{amssymb}
\usepackage{graphicx}

\makeatletter
\@ifundefined{textcolor}{}
{%
 \definecolor{BLACK}{gray}{0}
 \definecolor{WHITE}{gray}{1}
 \definecolor{RED}{rgb}{1,0,0}
 \definecolor{GREEN}{rgb}{0,1,0}
 \definecolor{BLUE}{rgb}{0,0,1}
 \definecolor{CYAN}{cmyk}{1,0,0,0}
 \definecolor{MAGENTA}{cmyk}{0,1,0,0}
 \definecolor{YELLOW}{cmyk}{0,0,1,0}
}

\makeatother

\usepackage{babel}
\begin{document}
\global\long\def\m{\mathrm{\,\mu m}}

\global\long\def\tg{T_{\mathrm{g}}}

\title{Capillary rupture of suspended polymer concentric rings}

\author{Zheng \surname{Zhang}}

\affiliation{Department of Mechanical Engineering, University of Colorado at Boulder,
Boulder, Colorado 80309, USA}

\author{G.~C. \surname{Hilton}}

\affiliation{National Institute of Standards and Technology, Boulder, Colorado
80305, USA}

\author{Ronggui \surname{Yang}}

\affiliation{Department of Mechanical Engineering, University of Colorado at Boulder,
Boulder, Colorado 80309, USA}

\author{Yifu \surname{Ding}}

\email[Correspondence: ]{yifu.ding@colorado.edu}

\affiliation{Department of Mechanical Engineering, University of Colorado at Boulder,
Boulder, Colorado 80309, USA}

\affiliation{Materials Science and Engineering Program, University of Colorado
at Boulder, Boulder, Colorado 80309, USA}

\date{\today}
\begin{abstract}
We present the first experimental study on the simultaneous capillary
instability amongst viscous concentric rings suspended atop an immiscible
medium. The rings ruptured upon annealing, with three types of phase
correlation between neighboring rings. In the case of weak substrate
confinement, the rings ruptured independently when they were sparsely
distanced, but via an out-of-phase mode when packed closer. If the
substrate confinement was strong, the rings would rupture via an in-phase
mode, resulting in radially aligned droplets. The concentric ring
geometry caused a competition between the phase correlation of neighboring
rings and the kinetically favorable wavelength, yielding an intriguing,
recursive surface pattern. This frustrated pattern formation behavior
was accounted for by a scaling analysis.
\end{abstract}


\maketitle
Capillary instability is a commonly observed phenomenon: a slender
liquid object ruptures into a series of droplets, driven by surface/interfacial
tension ($\gamma$) \citep{eggers2008physics}. The droplets are spaced
at a characteristic distance, corresponding to the fastest growing
wavelength (mode). From Tomotika's linear stability analysis, this
mode is a function of the interfacial tension and the cylinder-to-medium
viscosity ratio \citep{tomotika1935onthe}. When multiple cylinders
are embedded in parallel within the same medium, the dominant mode
for neighboring cylinders can become correlated \citep{elemans1997development}:
the droplets positioned either in-phase or out-of-phase \citep{knops2001simultaneous}.

Instability of non-minimal shapes is fundamentally interesting. However,
despite the rich literature on capillary instability of straight cylinders,
studies on curved objects have been rather lacking until recent years.
Pairam et al.\ \citep{pairam2009generation} successfully created
an unstable ring (toroid) by injecting liquid into a rotating bath
of an immiscible liquid. They showed that the evolution of the as-formed
ring was dictated by the competition between radial contraction and
circumferential rupture \citep{pairam2009generation}. Yao et al.\ analyzed
the Stokes flow during the contraction \citep{yao2011theshrinking}.
Mehrabian et al.\ simulated both the contraction and the non-linear
rupture of an embedded Newtonian ring \citep{mehrabian2013capillary}.
The aforementioned literature suggests that the characteristic contraction
time and rupture time predominantly scale with the medium viscosity
and ring viscosity, respectively. Indeed, by replacing the medium
with a highly viscoelastic material, the two timescales can be decoupled
\citep{pairam2014stability}. Furthermore, the stability of a substrate-supported
liquid ring was studied both theoretically \citep{gonzalez2013stability}
and experimentally, via spin-coating \citep{mcgraw2010plateaurayleigh},
solvent evaporation \citep{byun2008evaporative} and pulse-laser \citep{wu2010onthe,nguyen2012competition},
as well as ion-beam \citep{lian2006patterning}. 

Previous research has focused on a single ring; whether and how multiple
closely arranged rings would rupture remained unclear. This could
be because none of the literature methods were capable of creating
multiple embedded rings, with well-defined dimensions and physical
properties. In this Letter, we report the first experimental study
on capillary instability amongst suspended concentric rings.

\begin{figure}
\includegraphics[width=0.75\columnwidth]{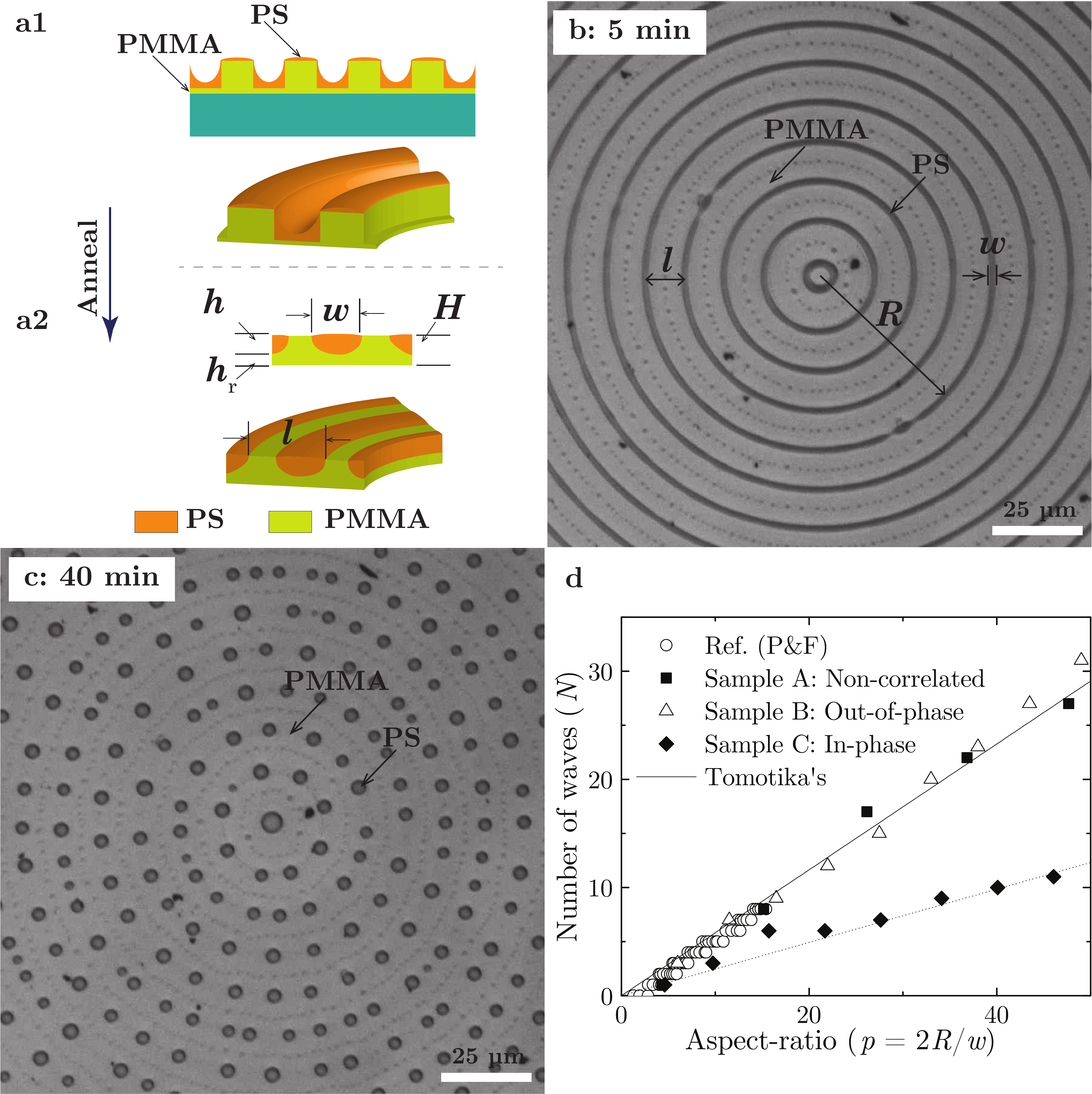}

\protect\caption{(a1-a2) Schematic of the formation of suspended PS rings upon annealing.
Optical image of concentric PS rings ($w=3.1\protect\m$, $l=20.2\protect\m$,
Sample A) (b) as formed and (c) after rupture, annealed at 160 \textdegree C
for the labeled durations. (d) The number of capillary waves plotted
as a function of aspect ratio. The solid line is based on Tomotika's
theory. The dotted line is a linear fit to the ``In-phase'' data.
The empty circles are adapted from reference \citep{pairam2009generation}.}

\label{fig1}
\end{figure}
The concentric rings were created by a three-step fabrication process,
which we developed previously \citep{zhang2012instabilities}. Briefly,
we first imprint a concentric ring pattern on a spin-coated poly(methyl
methacrylate) (PMMA) film, via nanoimprint lithography. Next, a layer
of polystyrene (PS) was spin-coated onto the PMMA pattern, using a
selective solvent (1-chloropentane).  The bilayer would form individual rings
upon annealing. In this Letter, we show three representative samples
(A, B and C), imprinted with different patterns.

The cross-sectional geometry of the as-cast patterns is illustrated
in Fig.~\ref{fig1}(a1), where PS mostly segregated in the PMMA trenches.
Being a non-minimum shape, the pattern would spontaneously evolve
at a temperature above the $\tg$s of PS and PMMA.

At first, the corrugation was leveled by the vertical Laplace pressure ($P\approx2\pi^{2}\gamma d/l^{2}$,
where $d$ and $l$ are the height and periodicity of the corrugation,
respectively \citep{ding2008nanoimprint,ding2010stability}), forming
PS rings atop PMMA. Hereafter we refer to the PS rings as the ``rings''.
Fig.~\ref{fig1}(a2) schematically shows the cross-section of the
 rings, with $w$, $h$, $h_{r}$ denoting the width and thickness
of the ring, and the residual layer thickness, respectively.

At 160 \textdegree C, the entire leveling process completed within
the first minute of annealing, as the flow times of both polymers
under $P$ were very short:  $\eta_{\mathrm{PS}}=802\,\mathrm{Pa\cdot s}$,
and $\eta_{\mathrm{PMMA}}=1450\,\mathrm{Pa\cdot s}$, from our rheological
measurements. During this process, $w$ of the rings decreased, in
order to balance the surface tension of PS and the interfacial tension
of PS/PMMA ($\gamma_{\mathrm{PS}}/\gamma_{\mathrm{PS/PMMA}}\approx24$)
\citep{zhang2012instabilities,mark2007physical}. The $R$ of each
ring and periodicity $l$ remained nearly constant, indicating negligible
contraction of the rings. The only exception was the innermost ring
in Fig.~\ref{fig1}b evolving into a single droplet in Fig.~\ref{fig1}c.
However, this was not due to radial contraction: the ring ruptured
first and then rounded up (Fig.~S2), which was also observed on fat
rings by Pairam et al.\ \citep{pairam2009generation} 

By adopting Yao \& Bowick's solution \citep{yao2011theshrinking},
the contraction rate for the 2\textsuperscript{nd} innermost ring
was estimated to be on the order of $10^{-3}\m\mathrm{/s}$. Before
the capillary rupture time of $\sim20\,\mathrm{min}$ (Fig.~S2c),
$R$ would only decrease $\sim1\m$, which was negligible in comparison
with $l$. The rest of the rings had even larger aspect-ratio ($p=2R/w$)
and, therefore, even smaller contraction rate. Here $w/2$ is considered
equivalent to the tube radius ($a$) of a toroid. In following discussions,
we will only focus on the rupture behaviors of the rings, after the
initial leveling process. 

Fig.~\ref{fig1}b shows Sample A at 5~min of annealing, forming
concentric rings with $w\approx3.1\m$, $h\approx1.0\m$, $h_{r}\approx1.1\m$
and $l\approx20.2\m$. All the rings had identical $w$, which guaranteed
$p\propto R$. The cross-section of each ring, represented by a ratio
of $w/h\approx3$, is consistent with that of straight filaments after
the fast leveling process \citep{zhang2012instabilities,ahn2010hierarchical,ahn2011vertical}.
At this time, the rings in the PMMA trenches remained continuous,
albeit periodic capillary fluctuations were already discernible (Fig.~\ref{fig1}b).
After 40~min, all rings had ruptured into discrete droplets (Fig.~\ref{fig1}c).
Note that the ruptured segments quickly equilibrated into droplets
and were kinetically immobilized, because collision-based coalescence
rate was extremely slow.

We plot the number of waves ($N$) from each ring as a function of
$p$ (solid squares in Fig.~\ref{fig1}d). The relationship agrees
well with Tomotika's theory for a cylinder (with a radius $a=w/2$)
embedded in an infinite medium (solid line in Fig.~\ref{fig1}d,
with a slope of 0.582)\citep{tomotika1935onthe,pairam2009generation}.
Since $p\propto R$, the linearity $N\propto p$ implies that the
breakup wavelength $\lambda=2\pi R/N\propto R/p$ was a constant for
all the rings with sufficiently large $p$. 

\begin{figure}
\includegraphics[width=0.75\columnwidth]{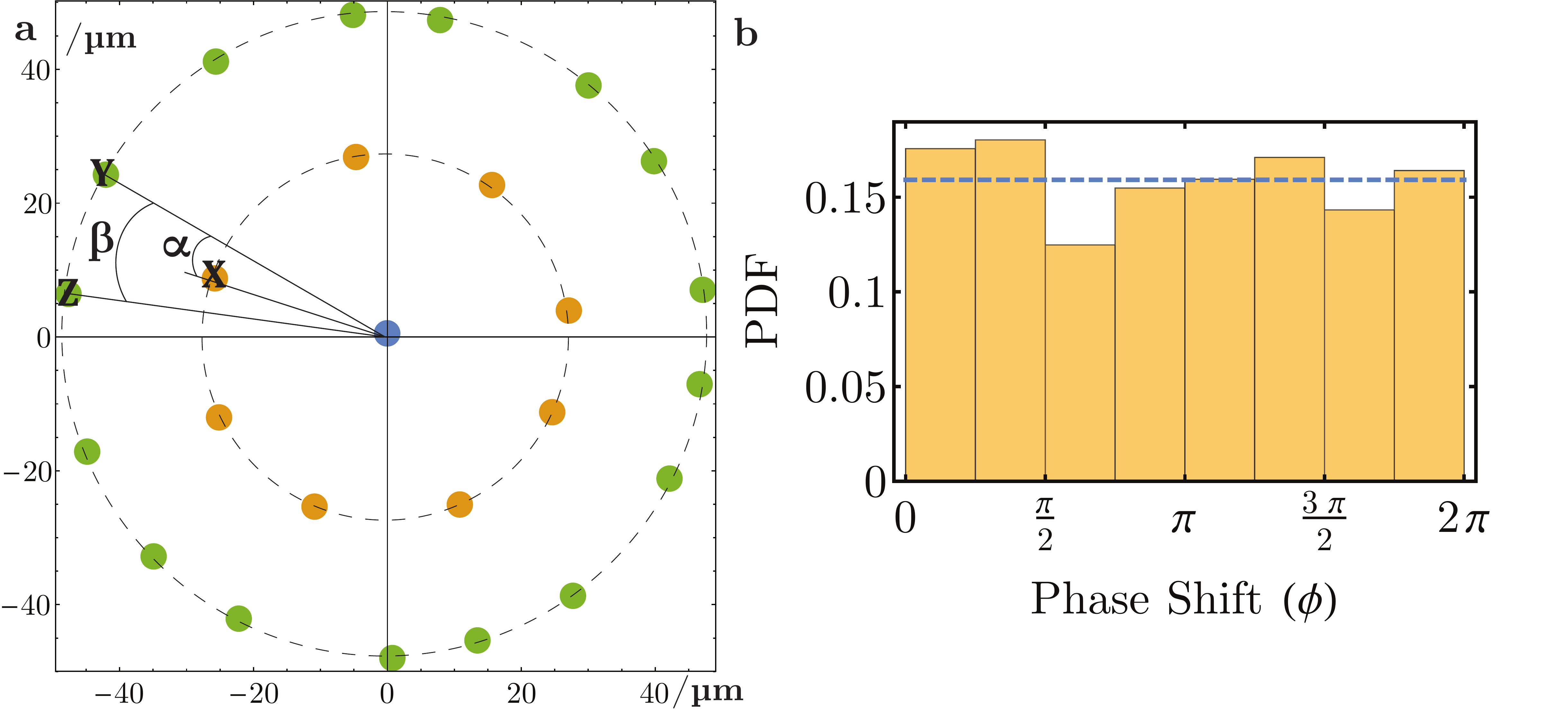}

\protect\caption{(a) Schematic for defining phase shift ($\phi$); (b) Distribution
of $\phi$. The dashed line is the mean average of the bar heights.}

\label{fig1_1}
\end{figure}
Besides $\lambda$, we were also interested in the phase correlation
between the rupture of neighboring rings. To unambiguously identify
the phase correlation, we statistical analyzed the coordinates of
all the droplets formed.We define the phase shift ($\phi$) \emph{locally}
for every droplet, as shown in Fig.~\ref{fig1_1}a. For an arbitrary
droplet X, we find its closest pair of droplets on the outer ring,
Y and Z, and calculate central angles $\alpha$ and $\beta$. We define
$\phi$ with an angular relationship: $\phi=2\pi\alpha/\beta.$

For in-phase correlation, X aligns with either Y or Z along the radus
(in-phase), so $\phi$ becomes $0$ or $2\pi$, respectively. For
out-of-phase, $\alpha=\beta/2$ and, therefore, $\phi=\pi$.  Fig.~\ref{fig1_1}f
is the distribution across the entire sample, suggesting a uniform
distribution: the rings ruptured independently. This is not surprising,
given that neighboring rings are sparsely distanced ($2l/w\approx13$).
Knops et al.\ showed that for a viscosity ratio of 0.04, the flow
induced by capillary rupture of a cylinder extended up to $\sim10$
times its radius \citep{knops2001simultaneous}.

\begin{figure}
\includegraphics[width=0.75\columnwidth]{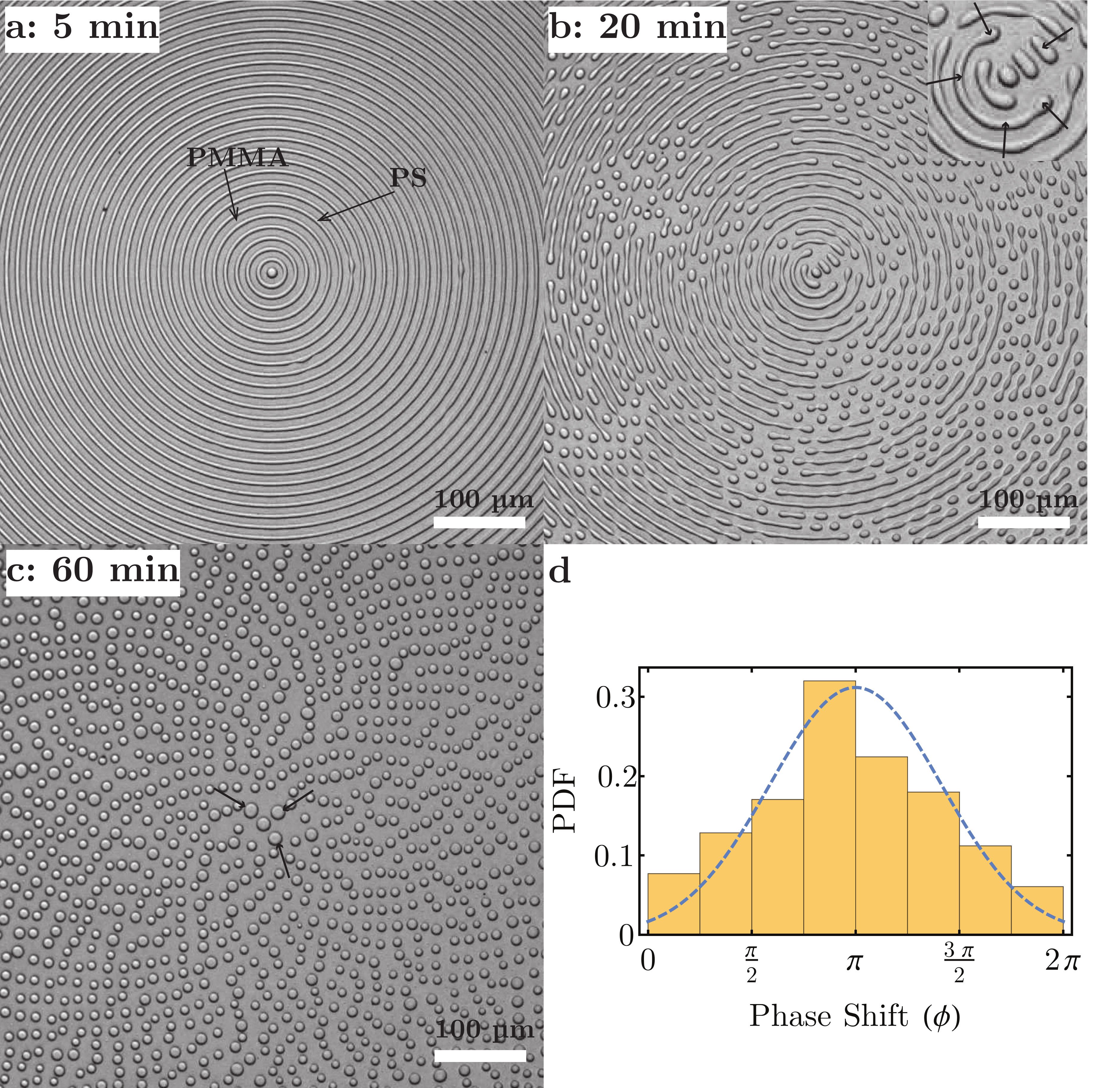}

\protect\caption{Optical images of Sample B annealed at 170 \textdegree C: (a) after
 leveling ($w=4.6\protect\m$, $l=12.1\protect\m$) and (b, c) rupture.
(d) Distribution of $\phi$. The dotted line is a fit to a truncated
normal distribution within domain $[0,\,2\pi)$.}

\label{fig2}
\end{figure}
In order to enhance the hydrodynamic interference between neighboring
rings, we fabricated Sample B with reduced $l$, via the same procedure
but with a different mold ($12\m$ periodicity, and a line-and-space
ratio of 1). Upon annealing at 170 \textdegree C for 5 min, the surface
leveling process was completed, resulting in a set of  denser packed
rings(Fig.~\ref{fig2}a) than Sample A (Fig.~\ref{fig1}b). For
Sample B: $w\approx4.6\m$, $h\approx1.4\m$, $h_{r}\approx1.7\m$
and $l\approx12.1\m$. The cross-section $w/h\approx2.7$, again,
reflected the balance between the $\gamma_{\mathrm{PS}}$ and $\gamma_{\mathrm{PS/PMMA}}$.
Most critically, the $2l/w$ ratio for Sample B was $\sim5.3$. Upon
further annealing, the rings started to undulate and rupture (Fig.~\ref{fig2}b).
After 60~min, all  rings had ruptured into discrete droplets (Fig.~\ref{fig2}c),
whose sizes and positions remained unchanged even after 540~min.
(Fig.~S3)

Similar as Sample A, the $N\sim p$ relationship (empty triangles
in Fig.~\ref{fig1}d) matched Tomotika's theory \citep{tomotika1935onthe,pairam2009generation}.
The only exceptions are the two innermost rings, where in-plane relaxation
dominated and reduced the number of droplets (arrows in Fig.~\ref{fig2}b
and Fig.~\ref{fig2}c.) 

In stark contrast, Fig.~\ref{fig2}d suggested a unimodal distribution,
peaked near $\pi$. This unambiguously shows that the most probable
phase correlation is out-of-phase. For this sample, the neighboring
rings were sufficiently near ($2l/w\approx5.3$) to interfere with
each other. Based on recent numerical work, for a cylinder/medium
viscosity ratio of $\sim1$, out-of-phase correlation is expected
for a $2l/w$ ratio of 3 -- 10) \citep{janssen2012stability}. Sample
A and B had a $2l/w$ ratio of 12 and 5.3, consistent with their non-correlated
or out-of-phase mode, respectively.

From the previous studies on \emph{straight filments} \citep{zhang2012instabilities,ahn2010hierarchical,elemans1997development,knops2001simultaneous}
, the out-of-phase mode is the result of synchronized flow amongst
neighbors: An alternation of necking and expanding occurred along
the orthogonal direction. Therefore, $N$ is constrained to be identical
between neighbors.  If this is also true for \emph{concentric rings},
it would contradict the observed $N\propto p$ (Fig.~\ref{fig1}d).
We owe the observed linearity to the locality of the out-of-phase
breakup, since there was no indication of long-range correlation/influence
across Sample B surface (Fig.~\ref{fig2}b and c). The correlation
became more evident starting from the 6\textsuperscript{th} ring
(Fig.~S4). This also resulted in a broadened distribution of $\phi$.

Further decreasing the spacing between neighbors could transition
the correlation into ``in-phase'', when the axial flow started to
couple amongst neighbors \citep{knops2001simultaneous,janssen2012stability}.
This was also observed in sheared polymer blends \citep{martys2001critical}.
From the recent numerical work \citep{janssen2012stability}, we expect
the threshold of $2l/w$ for out-of-phase to in-phase transition to
be $\sim3$, for our system (viscosity ratio of $\sim0.55$). However,
fabricating so densely-packed rings turned out rather challenging:
Simply increasing the cast volume of PS (higher concentration or low
spin speed) would only result in a thick top layer, which levels into
a planar bilayer during annealing, as opposed to forming concentric
rings \citep{zhang2012instabilities}.

We recently discovered that strongly confined straight filaments (e.g.
small $h_{r}$) always break up in-phase, regardless of the viscosity
ratio or the substrate wettability \citep{zhang2013influence}. Herein,
we fabricated substrate confined rings (Sample C). The degree of confinement
can be defined as $H/h$, where $H$ is the overall film thickness
(Fig.~\ref{fig1}a2). The smaller $H/h$ is, the stronger substrate
confinement is. The $H/h$ for Sample A (Fig.~\ref{fig1}) and B
(Fig.~\ref{fig2}) were 2.1 and 2.2, respectively; both were larger
than the bulk-to-confinement threshold of 2.0 \citep{hagedorn2004breakup}.
Therefore, both cases can be considered as weakly confined.

\begin{figure}
\includegraphics[width=0.75\columnwidth]{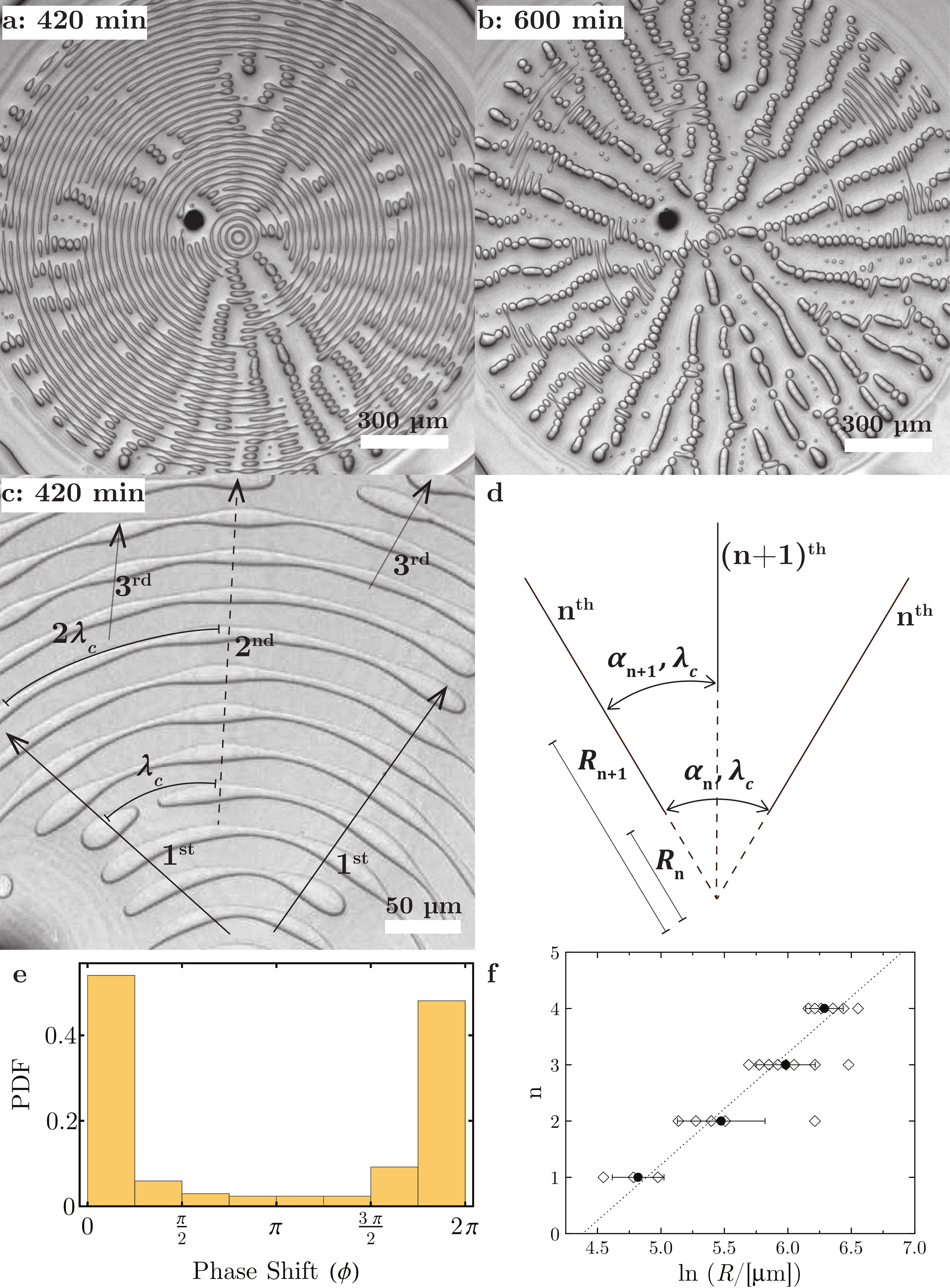}

\protect\caption{(a, b) Optical images of Sample C annealed at 170 \textdegree C for
the labeled durations. Before undulation,  $w=8.5\protect\m$, $l=25.0\protect\m$.
(c) Close-up view of the undulation. (d) Schematic for the recursively
``inserted'' waves. (e) Distribution of $\phi$. (f) Scaling between
the order of generation $n$ and $R$. Empty diamonds represent each
new wave. The dotted line is a linear fit.}

\label{fig3}
\end{figure}
Fig.~\ref{fig3} shows Sample C (see Fig.~S5 for more snapshots).
After the initial surface leveling within 5~min (and stable up to
180~min), $w\approx8.5\m$, $h\approx1.6\m$, $h_{r}\approx0.5\m$,
$l\approx25.0\m$. For Sample C, $H/h=1.2$, meaning substrate exerted
strong confinement on the rings. Its $w/h\approx5.3$, indicating
a flattened ribbon shape, that deviated significantly from the equilibrium
shape ($w/h\approx3$) of a weakly confined thread.

These confined rings were much more kinetically stable: they started
to rupture between 400--600 min, which was more than one order of
magnitude slower than Sample A and B (Fig.~\ref{fig3}). The difference
cannot be adequately explained by their difference in $w$. This is
consistent with literature showingsuppressed capillary instability
under confinement \citep{migler2001stringformation,martys2001critical,son2003suppression}.
Recent work by Alvine et al.\ also showed that the capillary fluctuations
of polymer melt were dramatically hindered atop a topographic Si grating
\citep{alvine2012capillary}.

Despite the slow kinetics, the rings eventually ruptured (Fig.~\ref{fig3}b).
However, different from straight filament arrays \citep{zhang2012instabilities,zhang2013influence},
these droplets radially lined up. We plot the $N\sim p$ scaling (diamond
symbols) in Fig.~\ref{fig1}d. The linearity again indicates constant
$\lambda$ for all rings. A linear fit (dotted line) shows that the
slope (0.25) is less than half of Tomotika's theory. This directly
translates to larger wavelength and droplet size by volume conservation. 

As previously discussed, for concentric rings, $N\propto p$ (also
observed in two other strongly confined patterns, Fig.~S6). Apparently,
more waves had been \textquotedblleft inserted\textquotedblright{}
into the outer rings. Here we attempt to shed light upon this process.
 Fig.~\ref{fig3}(c) shows the undulation. The primary correlated
directions are marked \textquotedblleft 1\textsuperscript{st}(generation)\textquotedblright ,
extending radially from the center and perpendicular to the tangential
of the rings. Moving away from the center, more waves were ``inserted''
in between the primary directions. Although the PS segments enveloped
between ``1\textsuperscript{st}'' directions all had the same central
angle, their arc length (also $\lambda$) increased with $R$. The
increased $\lambda$ required a gradually less favorable undulation
mode and built up the level of frustration. When this frustration
grew sufficiently large, it could be released by inserting an additional
wave in between (marked ``2\textsuperscript{nd}'' in Fig.~\ref{fig3}c).
Similarly, the 3\textsuperscript{rd} generation can be found at an
even larger $R$.

Therefore, the most energetically favorable (least amount of frustration)
mode should correspond to the smallest $\lambda$. We denote this
characteristic wavelength with $\lambda_{c}$. The upper bound of
$\lambda$ should be the most frustrated wavelength $2\lambda_{c}$
(on the verge of splitting up into two waves). Thus we have $\lambda_{c}<\lambda<2\lambda_{c}$.
$\lambda_{c}$ can be directly measured by identifying the smallest
wavelength, as labeled in Fig.~\ref{fig3}(c). We obtained that $\lambda_{c}=77.8\pm11.0\m$
and the average $\lambda=108.9\pm7.0\m$, which was consistent with
the lower and upper bound limit. Tomotika's theory provides a wavelength
estimate of $45.9\m$. However, the lower bound $\lambda_{c}$ was
larger than the prediction, due to confinement-induced wavelength
increase \citep{son2003suppression,hagedorn2004breakup}. 

We develop a scaling relationship to capture the recursive nature
of the ``insertion'' behavior. As shown in Fig.~\ref{fig3}d, in
between the $n$\textsuperscript{th} generation envelope (radius
$R_{n}$, central angle $\alpha_{n}$, wavelength $\lambda_{c}$),
a new wave is inserted in the middle but at a larger radius $R_{n+1}$.
Therefore, $\forall n\in\mathbb{N}$:
\begin{align*}
\alpha_{n}=\frac{\alpha_{1}}{2^{n-1}},\  & R_{n}\alpha_{n}=\lambda_{c}\Longrightarrow R_{n}=\frac{\lambda_{c}}{2\alpha_{1}}\cdot2^{n}\propto2^{n},
\end{align*}
or equivalently $\ln R_{n}\propto n$. We statistically verify the
scaling against the experimentally observed order of generation $n$
and $\ln R_{n}$ (Fig.~\ref{fig3}f and Supplementary Fig.~S7). The
dotted line is a fit to the mean value of $\ln R_{n}$ for each generation
($n$). The excellent linearity proves that, despite the randomness
of surface capillary instability, our simple scaling analysis was
capable of capturing the essential ``recursive'' behavior of the
concentric rings.

In summary, we developed a novel procedure that allowed us to examine
capillary breakup of concentrically arranged PS rings, suspended atop
a layer of PMMA. When the substrate confinement was weak, the rings
broke up independently if they were far apart, but via an out-of-phase
mode if they were sufficiently close. For both cases, the breakup
wavelength agreed well with the prediction by Tomotika's linear stability
theory for a fully embedded cylinder (approximating the ring half-width
as the cylinder radius). Under significant confinement of the substrate,
the rings tended to breakup via an ``in-phase'' mode along the radial
direction. The unique concentric ring geometry induced strong geometric
frustration, which yielded a self-similar morphology that could be
accounted for by our scaling analysis. Geometric frustration associated
with curvature is a fundamentally important topic. Our experiments
can serve as a basis for correlated capillary instability among curved
objects, which can be a powerful tool for creating unique surface
patterns.
\begin{acknowledgments}
\appendix
This work was supported by the National Science Foundation under Grant
CMMI-1031785 and CMMI-1233626. ZZ acknowledges support from the Beverly
Sears Graduate Student Grant at CU-Boulder.
\end{acknowledgments}

\bibliography{concentricrings}

\end{document}